# Healthcare Utilization and Perceived Health Status from Falun Gong Practitioners in Taiwan: A Pilot SF-36 Survey

Yu-Whuei Hu[1,2], Ph.D., Li-Shan Huang[3,4], Ph.D., Eric J. Yeh[5], Ph.D., Mai He[6*], M.D., Ph.D

1 Department of Economics, National Dong Hwa University, Hualian, Taiwan

2 Department of Economics, National Taiwan University, Taipei, Taiwan

3 Department of Biostatistics and Computational Biology, University of Rochester, Rochester, NY, 14642, USA

4 Institute of Statistics, National Tsing Hua University, Hsinchu, 30013, Taiwan

5 Amgen Inc. Thousand Oaks, CA91320, USA

6 Department of Pathology & Immunology, Washington University School of Medicine, St. Louis, MO 63110, USA

*Corresponding author:

Mai He, M.D., Ph.D.

Associate Professor

Department of Pathology & Immunology

Washington University School of Medicine

660 South Euclid, Campus Box 8118

St. Louis, MO 63110

Phone: 314-273-1328

Email: Maihe@wustl.edu

**Abstract**

**Objective:** Falun Gong (FLG) is a practice of mind and body focusing on moral character improvement along with meditative exercises. This 2002 pilot study explored perceived health status, medical resource utilization and related factors among Taiwanese FLG practitioners, compared to the general Taiwanese norm estimated by the 2001 National Health Interview Survey (NHIS).

**Methods:** This cross-sectional, observational study was based on a voluntary, paper-based survey conducted from October 2002 to February 2003 using the same Taiwanese SF-36 instrument employed by the NHIS. Primary outcomes included eight SF-36 domain scores and the number of medical visits. One-sample t-tests, one-way ANOVA and multivariate linear regression analyses were performed.

**Results:** The response rate was 75·6% (1,210/1,600). Compared to the norm, the study cohort had significantly higher scores in six of eight SF-36 domains across gender and age ($p<0\cdot05$). Among those with chronic diseases, 70% to 89% reported their conditions either improved or cured. 74.2%, 79.2%, 83.3%, and 85.6% quitted alcohol drinking, smoking, chewing betel nuts, and gambling. 62·7% reported a reduced number of medical visits (mean=13·53 before; mean=5·87 after).

**Conclusions:** In this subject cohort, practicing FLG led to higher perceived health scores and reduced health resource utilization compared to the norm.

utilization

## Human Subject Approval Statement

At the time of the study period (October 2002 to February 2003), IRB approval was not required to conduct a questionnaire survey based on the guidelines of National Taiwan University where the PI, Dr. Yu-Whuei Hu, was affiliated. The dataset was de-identified, and the PI, the holder of the key, deceased in 2006. Subsequent use of de-identified data would not constitute human subjects research, since it is no longer identifiable. The data are exempt under 32 CFR 219.101(b)(4).

## Informed Consent Requirement

Not applicable.

## Conflict of Interest Disclosure Statement

All authors claim no conflict of interest

**Funding:** None.

## Introduction

It is well recognized that escalating health expenditure is a global challenge.[1,2] In response, health care has evolved into an outcome-based, patient-centered, integrated care model. There is an increasing demand among various (integrated) healthcare models to include disease prevention and wellness programs.[3] An important component of preventive and wellness programs is to integrate effective complementary and alternative approaches such as diet, nutrition, exercise and patient self-management of health to improve wellbeing.[4]

Yoga, Tai Chi and qigong have received increased attention lately. National surveys indicate that increasing number of adults in the U.S. practice yoga, Tai Chi or qigong (6·7% in 2002, 7·4% in 2007 and 10·9% in 2012, respectively).[5] The relationships of yoga, Tai Chi or qigong with health benefits have been evaluated.[6–9] Among available complementary health approaches, Falun Gong has been over-looked in the scientific literature.

As a type of qigong, Falun Gong, (also known as Falun Dafa), was introduced to the public in 1992 in China, and has since spread to countries around the globe, including Taiwan and the United States.[10] Falun Gong is similar to regular qigong, yoga, and Tai Chi with slow-motion exercises and sitting meditation, but differs with its emphasis on mental and moral character improvement by following the core principles of "Truthfulness, Compassion and Tolerance." Reading the book *Zhuan Falun*[11] and applying these principles in daily life is regarded as the main approach on spiritual improvement.

The estimated number of those who were practicing Falun Gong (practitioners or adherents) in China increased to almost 70 million between mid-1992 to 1999,[12] with growth partially due to the effect or belief that practicing Falun Gong would improve health and reduce medical costs.[13] Although banned in July 1999, China continues to have the largest number of Falun Gong practitioners, followed by Taiwan.[10] Thus, this second largest pool of Falun Gong practitioners in Taiwan is the best available source for this study.

This pilot survey from October 2002 to February 2003 aimed to investigate the effect of practicing Falun Gong on perceived health status and healthcare visits. The objectives of this pilot study were (1) to describe characteristics, perceived health status and medical visits before and after practice among Falun Gong practitioners in Taiwan, (2) to compare (1) with the norm in the general public, and (3) to explore key factors that would influence health status and medical visits among Falun Gong practitioners.

Since the pilot study was performed about 15 years ago, we recognize that it may not be representative of the population of Falun Gong practitioners of Taiwan today. Nevertheless, the year 2002 was close to the 2001 National Health Interview Survey (NHIS) conducted in Taiwan and both studies employed the same SF-36[14,15] questionnaire. It is thus possible to make a comparison between the general population and Falun Gong practitioners in Taiwan at nearly the same time. In addition, given that there is no published medical literature on the health status and health care utilization of Falun Gong at the community level, the results of this pilot study from 2002 will be of interest to a wide range of health care communities and, considering the challenges in today's healthcare systems, may help assess the feasibility of future full-scale surveys.

## Materials and Methods

### *Study design and Participants*

This pilot study employed an observational and cross-sectional design. The target population was Falun Gong practitioners (adherents) in Taiwan. Inclusion criteria of study subjects were (1) practice of Falun Gong for more than two months, (2) beginning reading the book *Zhuan Falun*[11], and (3) at least 18 years of age at the time of survey. Exclusion criteria were (1) less than 18 years of age, (2) unable to read and respond to the survey in Chinese, and (3) unwillingness to participation.

A multistage stratified randomized cluster sampling method was employed in an attempt to maximize representation. Based on the degree of development of the cities and towns, 350 townships of Taiwan were stratified into 10 levels. A cluster was formed by randomly selecting 20 percent of the cities/towns within each level. Falun Gong practitioners in the chosen towns were solicited to participate in this survey. The cohort consisted of voluntary participants answering a paper-based survey that was implemented during the study period (from October 2002 to February 2003) to obtain participants' demographic information and self-assessment of health status.

At the time of the study period (October 2002 to February 2003), IRB approval was not required to conduct a questionnaire survey based on the guidelines of National Taiwan University where the PI, Dr. Yu-Whuei Hu, was affiliated. The dataset was de-identified, and the PI, the holder of the key, deceased in 2006. Subsequent use of de-identified data would not constitute human subjects research, since it is no longer identifiable. The data are exempt under 32 CFR 219.101(b)(4).

***Survey Instrument: Short form – 36 (SF-36) Health Survey***

The SF-36 health survey consists of 36 multiple operational indicators measuring physical and mental components of health status across eight domains.[14] The physical component is measured by four domains: physical function (PF), role limitation due to physical problems (RP), bodily pain (BP) and general health (GH). The mental component is measured by four domains: vitality (VT), social function (SF), role limitation due to emotional problems (RE) and mental health (MH). To allow comparison with the general population in Taiwan, this survey adopted the same SF-36 health survey used in Taiwan's NHIS conducted in 2001.[15] The validation process of the Taiwan version of SF-36 was described elsewhere.[16] All the eight domain scores were taken as the primary outcomes in our study.

The other primary outcome was health resources utilization, measured by the frequency of medical visits, which was obtained from the record on the practitioners' health insurance cards issued by the Bureau of National Health Insurance in Taiwan at the time. In addition to primary outcomes, the survey also collected demographic information, socioeconomic status, health related information and Falun Gong practice related information (see Appendix 1, in Chinese).

***Power analysis***

Please see Appendix 2 for details. To detect a 2-point change using a one-sample t-test with two-sided alpha of 0·05, 1,000 subjects will yield approximately statistical power of 82%.

***Survey Administration & Data Collection***

Paper copies of the study questionnaire were sent to contact persons of practice sites in the selected towns as described earlier. Completed and returned written questionnaires were transferred into a digital format and numerically coded for analysis.

***Data Analysis***

Details of data analysis is described in Appendix 2. Briefly, the survey data were analyzed to examine the effect of practicing Falun Gong in the following four categories: (1) comparison of perceived health status to the 2001 Taiwan norm using one-sample t-test, (2) association of independent variables with perceived health status using one-way analysis of variance (ANOVA) and multivariate linear regression analyses, (3) association of independent variables with the number of medical visits by multivariate linear regression, and (4) testing interaction effect due to prior chronic conditions in multiple linear regression models by F-tests.

## Results

### *Univariate Analysis*

Sixteen hundred questionnaires were sent out and 1,210 were completed and returned; the response rate was 75·6%. Table 1 shows demographic characteristics of the study subjects and the estimated norm of the general population in Taiwan. Compared to the 2001 norm, the study subjects had a higher estimate in the following demographic and prior-practice health characteristics: female (59·8% vs. 49%,), the proportion of elderly (65 years and older, 14·9% vs. 9%), college education or above (36·1% vs. 21%), married (77·7% vs. 56%), monthly income (NTD $56,200 vs NTD $33,500), cardiovascular diseases (19·5% vs. 6·8%), diabetes (6·4% vs. 4·5%), pulmonary disorders (18·5% vs. 5·2%) and hypertension (17·0% vs. 11·3%), whereas a smaller estimate in private sector workers (39·0·% vs. 62 %) and students (6·1% vs. 29%).

Table 2 shows perceived health benefits, indicated by the perceived change in prior chronic condition(s) and the change in medical visits per year after practicing Falun Gong among study subjects. Cardiovascular diseases, diabetes, pulmonary disorders and hypertension were four of top chronic conditions in 2001 in Taiwan. Among the study subjects who had any of these four chronic conditions before practicing Falun Gong, the vast majority (ranging from 70% to 89%) reported that their health status was either cured or improved after practicing Falun Gong. A general trend of decreased numbers of medical visits was observed. Among all study subjects, the average number of medical visits per year before practicing Falun Gong was 11·96, which was similar to the previously reported mean in the general population in Taiwan during the period of 1996-2001 [mean=14·4].[17] After practicing Falun Gong, the number of medical visits per year dropped to 5·93. Excluding 59 subjects with missing values, 62·7% of subjects reported a reduced number of medical visits, indicating a drop in medical resource utilization in the study cohort, while 30·9% reported the same number of medical visit before and after practice.

For the behavior question on alcohol drinking, 213 people answered yes to "have a desire to drink" before practicing Falun Gong, and 158 (74.2%) of those quitted drinking after practice. Similarly, for smoking, chewing betel nuts, and gambling, 79.2%, 83.3%, and 85.6% quitted out of 125, 42, and 111 people respectively.

***Bivariate Analysis***

The perceived health benefits measured by SF-36 instrument were also observed. Table 3 shows the comparisons of the SF-36 mean domain scores between subjects and the general population in Taiwan by gender and age, respectively. Overall, study subjects had a significant higher mean SF-36 domain scores than those in the norm in Taiwan in six out of eight health dimensions ($p <= 0·001$). Non-significant results were observed within PF and SF domains. When survey results were analyzed by gender, similar patterns of comparisons were observed among males and females except for the mean PF score in females. Similar patterns were again observed in the age 40-64 and 20-39 groups. Among age 65+, all of the mean SF-36 domain scores were significantly higher in the surveyed Falun Gong practitioners than in the general population ($p<0·001$).

Table 4 shows the effect of the length of practicing Falun Gong on each SF-36 domain score among study subjects. Generally, the SF-36 domain scores increase with an increased length of practicing Falun Gong with few exceptions. One-way ANOVA indicated that there were significant differences in SF-36 domain scores among subjects with different length of practice ($p<0·001$ in six out of eight domains, except for RP and RE). These results suggested the length of practicing Falun Gong had a positive impact on perceived health status.

***Multivariate Analysis***

Table 5 shows key results of multivariate linear regression analyses where each column shows results of one regression analysis. The R-squared ranging from 0·0735 to 0·2621 indicates that there were other factors influencing the dependent variables and were not included in the regression models.

Based on results in Table 5, the number of times of reading the book *Zhuan Falun*[11] had the strongest positive association on all SF-36 domains and a significantly negative association on the number of medical visits, when other covariates (socioeconomic characteristics and prior chronic conditions) were controlled. The length of practice also had a significantly positive association on BP, GH, VT.SF and MH domain scores and had a negative association on the number of medical visits. The main effect of prior chronic conditions was significantly negative in all SF-36 outcomes except RE, and led to an increased number of medical visits. Higher household income was associated with better perceived health status.

***The interaction effect due to prior chronic conditions***

We tested if the effects of indeoendent variables (IVs) on primary outcomes were different between those who had prior chronic conditions and those who did not. The resulting F-tests were not significant for all nine linear models.

## DISCUSSION

While studies have shown an increase in the use of complementary and alternative therapies, [5,6,7] studies of the effect of Tai Chi or qigong on health care utilizations seem scarse.[18] This pilot study was the first research that investigated the effect of qigong on health care utilization besides the health status among Falun Gong practitioners in Taiwan in comparison with the norm in the Taiwanese general population at the time when this study was conducted. Very little can be found in the medical literature about Falun Gong practice.[19,20] The current study provided preliminary evidence on the association between Falun Gong practice and health benefits (SF36 scores and perceived changes on chronic conditions), and the potential impact on health care burden (number of medical visits).

***Health benefits***

In summary, results of this pilot SF-36 health survey demonstrated positive perceived health benefits and reduced medical visits from practicing Falun Gong in Taiwan when compared to the 2001 SF-36 health survey of the general Taiwanese public. Key findings included (1) on average, the Falun Gong practitioners were feeling healthier than the general population with fewer medical visits; (2) positive perceived health benefits of practicing Falun Gong were particular prominent among elderly people and those with chronic medical disorders; (3) effect of reading *Zhuan Falun* was significant in improving perceived health and reducing the number of medical visits; (4) an increased length of practicing Falun Gong improved perceived health status and reduced medical visits; and (5) interaction effect due to prior chronic conditions was unlikely in the perceived health benefits of practicing Falun Gong.

**The unique features of Falun Gong**

Yoga, Tai Chi and qigong including Falun Gong practice have similarities in that all these practices aim to cultivate a healthy body through slow-motion exercises and siting meditation. Falun Gong stands out from other disciplines as it does not require specific breathing techniques and it places strong emphasis on the improvement of moral character and changes in perception of stressful events. Therefore, Falun Gong is the only integrated system that cultivates a healthy body, mind and spirit.[10]

Based on the guiding principles of "Truth, Compassion and Tolerance", Falun Gong requires both practice (i.e., exercise and meditation) and cultivation (i.e., mind and spirit). Its philosophic principles are explained in the teachings of *Zhuan Falun*[11]. Frequent study of Zhuan Falun and application of the guiding principles in daily life are regarded as the most important part of the discipline. The strong association between reading this book with SF-36 scores and reduced medical visits supports the emphasis of cultivating one's mind and thoughts and the development of moral characters.

Based on these unique features, Falun Gong could be regarded as a self-practice, wellness-improving system integrating/consolidating physical (body), emotional (mind), social and spiritual wellbeing at multiple levels. These features fit into the current understanding of multiple aspects of health as defined by WHO "Health is a state of complete physical, mental and social well-being and not merely the absence of disease or infirmity". [21]

**Potential factors or mechanisms for the health promotion effects of practicing Falun Gong**

The potential mechanism of qigong is a difficult topic due to its complex nature. The concept of qi energy generation is hard to accept by the current evidence-based medical system. Li and colleagues investigated the impact of Falun Gong practice on gene expression profiles of neutrophils at cellular and molecular levels. [20] Neutrophils from Falun Gong practitioners have a significantly longer life span *in vitro* and increased functionality compared to cells from normal healthy controls. The comparison demonstrated enhanced immunity, down-regulation of cellular metabolism and more rapid resolution of inflammatory responses among Falun Gong practitioners than the controls. Review of gene expression studies of mind-body interactions revealed changes opposite to the effects of chronic stress.[22] Similarly, a systematic review suggested that Tai Chi practice boosts cell-mediated immunity and antibody response in immune

system. [23]

This pilot study had several limitations. (1) Due to the nature of a cross-sectional survey, causality cannot be assumed. Survey results may only be applicable to a certain population at a certain period of time. (2) The SF-36 outcomes were subject to recall bias, yet this bias was less likely for the frequency of medical visits, as the health insurance cards at the time carried this information. (3) The sampling frame was of Falun Gong practitioners in Taiwan, and hence the results may not be generalized to other populations in other countries. The sampling approach included self-selection, which may not yield a representative sample, but the use of clusters and the high response rate increased the likelihood of adequate representation. (4) The use of statistics in the general population (the "norm") may be viewed as an unfair comparison. However, our study approach was similar to that in the study conducted by Komelski and colleagues who compared health status between a sample of taijiquan and qigong practitioners to the U.S. representative sample.[24] Their results also indicated health status of taijiquan and qigong practitioners was significantly better than the US norm. We believed use of the norm in the comparison was reasonable. (5) This pilot study did not compare health benefits among those with four chronic conditions at baseline between the study cohort and the general public. Patients with prior chronic conditions may behave differently than those without any prior conditions. Our analysis suggests such interaction effect was less likely. Nevertheless, further research may be needed to investigate the effect of Falun Gong practice for particular diseases or conditions.

In conclusion, this pilot study demonstrates health benefits of practicing Falun Gong as manifested by improved perceived health status, as demonstrated in patient and outcome based measurements, with reduced medical visits, suggesting an enhanced cost-effectiveness of the practice to current health care. This study suggests the value of further research on the effects of this discipline. The significant association of frequent *Zhuan Falun*[11] study supports the emphasis on moral characters improvement of this qigong style and it deserves further studies.

## ACKNOWLEDGEMENTS

This manuscript was submitted in memory of late Dr. Hu who was the initial principal investigator of this study. The authors thank Prof. Meng-Chi Lio (National Sun Yat-Sen University, Kaohsiung, Taiwan) who helped develop the socioeconomic part of the survey, and Ann F. Corson, MD, for comments, grammar and formatting checks. And thanks to the survey participants who have willingly shared their precious time to make this project possible.

**Table 1**. Characteristics of study subjects and the general population in Taiwan

| % or mean (SD) | Falun Gong Practitioners | Norm* | p-value |
|---|---|---|---|
| N | 1210 | 17,515 | |
| Female (n=1210) | 59·8% | 49% | < ·0001 |
| Age, years (n=1208) | 49.8 (13·5) | | |
| Age, >=65 years | 14·9% | 9% | <·0001 |
| Sampling unit (n=1209) | | | ·0009 |
| City | 31·8% | 31% | |
| Town | 37·3% | 42% | |
| Others | 30·9% | 28% | |
| Household income (NTD/month)(n=1084) | $56,200 ($38,600) | $33,500 | <·0001 |
| Education level (n=1202) | | | <·0001 |
| Junior High | 13·0% | 24% | |
| Senior High | 26·5% | 28% | |
| College or above | 36·1% | 21% | |
| Others | 24·4% | - | |
| Marital Status (n=1205) | | | <·0001 |
| Married | 77·7% | 56% | |
| Divorced, widowed or separation | 9·5% | 10% | |
| Others | 12·9% | 34% | |
| Children (n=1155) | | | |
| No child | 14·9% | - | |
| The youngest child's age < 6 | 7·4% | - | |
| The youngest child's age 6-18 | 26·1% | - | |
| The youngest child's age ≥18 | 51·5% | - | |
| Employment, of workers (n=608) | | | <·0001 |
| Employer | 6·3% | 5% | |
| Own-Family work | 23·0% | 23% | |
| Government | 31·7% | 10% | |
| Private | 39·0% | 62% | |
| Of unemployed people (n=592) | | | <·0001 |
| Students | 6·1% | 29% | |
| Housekeeper | 54·6% | 36% | |
| Retiree | 29·9% | 25% | |
| Others | 9·5% | 10% | |
| Medical conditions | Before/after practicing Falun Gong | | |
| Cardiovascular diseases (n=1195) | 19·5%/12·6% | 6·8% | <·0001/<·0001 |
| Diabetes mellitus (n=1198) | 6·4%/4·9% | 4·5% | ·0016/·5224 |
| Pulmonary disorders (n=1194) | 18·5%/10· 6% | 5·2% | <·0001/<·0001 |
| Hypertension (n=1180) | 17·0%/10·4% | 11·3% | <·0001/·3656 |
| Non-Smoking (n=1210) | 97·8% | - | |
| Non-Alcohol drinking (n=1210) | 94·3% | - | |
| Length of practicing Falun Gong (n=1201) | | | |
| <½ year | 12·8% | - | |
| ½ to 1 year | 14·8% | - | |
| 1 to 2 years | 30·1% | - | |
| >2 years | 42·3% | - | |
| Times of reading *Zhuan Falun* (n=1199) | | | |
| < 1 time | 4·2% | | |
| 1 to 4 times | 21·4% | - | |
| 4 to 7 times | 12·3% | - | |
| 7 to 10 times | 7·9% | - | |
| More than 10 times | 54·1% | - | |

* Based on data from the 2001 National Health Interview Survey, Taiwan (ref 16)
Sum of percentages may not equal 100% due to rounding

**Table 2**. Perceived health benefits after practicing Falun Gong

| Among those with prior chronic conditions before practice | | | | |
|---|---|---|---|---|
| Status of chronic conditions before practice | Serous or very serious | | Mild | |
| Status of chronic conditions after practice | Cured | Improved | Cured | Improved |
| Cardiovascular diseases | 15/37 (41%) | 19/37 (51%) | 67/196 (34%) | 106/196 (54%) |
| Diabetes mellitus | 2/17 (12%) | 12/17 (71%) | 15/60 (25%) | 25/60 (42%) |
| Pulmonary disorders | 26/56 (46%) | 26/56 (46%) | 68/165 (41%) | 69/165 (42%) |
| Hypertension | 17/42 (41%) | 19/42 (45%) | 61/159 (38%) | 72/159 (45%) |

| Frequency of medical visits among all subjects | | |
|---|---|---|
| Medical visits per year | Before practicing Falun Gong | After practicing Falun Gong |
| None | 6·69% | 38·10% |
| 1-6 | 32·98% | 33·39% |
| 7-12 | 18·68% | 10·99% |
| 13-18 | 12·56% | 7·44% |
| 19-24 | 7·77% | 4·63% |
| 25-30 | 9·09% | 1·82% |
| 31-36 | 7·60% | 3·39% |
| No response | 4·63% | 0·25% |

**Table 3.** SF-36 domain scores of study subjects and the Taiwan general population by gender and age group

| Domain | Score | *PF* | *RP* | *BP* | *GH* | *VT* | *SF* | *RE* | *MH* |
|---|---|---|---|---|---|---|---|---|---|
| **Total** | | | | | | | | | |
| Falun Gong Practitioners | N | 1204 | 1198 | 1202 | 1189 | 1197 | 1205 | 1186 | 1191 |
| | Mean | 92·82 | 89·55 | 81·21 | 81·22 | 74·35 | 86·98 | 89·85 | 78·04 |
| | SD | (13·43) | (24·42) | (18·70) | (18·65) | (17·21) | (14·01) | (25·38) | (15·51) |
| Norm* | Mean | 92·21 | 83·61 | 73·33 | 70·42 | 68·25 | 86·80 | 79·40 | 73·01 |
| | SD | (16·20) | (33·30) | (28·57) | (21·78) | (18·67) | (17·07) | (36·07) | (16·55) |
| Difference | Mean | **0·61** | **5·94** | **7·88** | **10·80** | **6·10** | **0·18** | **10·45** | **5·03** |
| | p-value | 0·114 | <0·001 | <0·001 | <0·001 | <0·001 | 0·653 | <0·001 | <0·001 |
| **Male** | | | | | | | | | |
| Falun Gong Practitioners | N | 484 | 481 | 483 | 478 | 481 | 485 | 476 | 478 |
| | Mean | 93·41 | 89·36 | 82·91 | 80·74 | 75·24 | 87·81 | 89·85 | 79·18 |
| | SD | (12·67) | (25·37) | (17·73) | (18·15) | (15·75) | (13·47) | (25·54) | (14·43) |
| Norm* | Mean | 93·97 | 86·41 | 77·05 | 72·74 | 70·94 | 87·86 | 81·25 | 75·08 |
| | SD | (14·66) | (30·83) | (28·11) | (20·72) | (17·90) | (16·56) | (34·64) | (15·95) |
| Difference | Mean | **-0·56** | **2·95** | **5·86** | **8·00** | **4·30** | **-0·05** | **8·60** | **4·10** |
| | p-value | 0·336 | 0·011 | <0·001 | <0·001 | <0·001 | 0·934 | <0·001 | <0·001 |
| **Female** | | | | | | | | | |
| Falun Gong Practitioners | N | 720 | 717 | 719 | 711 | 716 | 720 | 710 | 713 |
| | Mean | 92·42 | 89·67 | 80·06 | 81·54 | 73·75 | 86·42 | 89·86 | 77·28 |
| | SD | (13·92) | (23·78) | (19·25) | (18·99) | (18·10) | (14·34) | (25·30) | (16·15) |
| Norm* | Mean | 90·49 | 80·87 | 69·67 | 68·14 | 65·61 | 85·76 | 77·60 | 70·97 |
| | SD | (17·40) | (35·35) | (28·56) | (22·54) | (19·04) | (17·49) | (37·33) | (16·88) |
| Difference | Mean | **1·93** | **8·80** | **10·39** | **13·40** | **8·14** | **0·66** | **12·26** | **6·31** |
| | p-value | <0·001 | <0·001 | <0·001 | <0·001 | <0·001 | 0·215 | <0·001 | <0·001 |
| **Age 20-39** | | | | | | | | | |
| Falun Gong Practitioners | N | 255 | 255 | 254 | 252 | 254 | 254 | 253 | 254 |
| | Mean | 97·47 | 94·80 | 83·73 | 84·85 | 73·23 | 87·11 | 88·93 | 75·91 |
| | SD | (7·95)[a] | (17·18) | (18·79) | (17·64) | (17·07) | (15·03) | (25·55) | (16·05) |
| Norm* | Mean | 96·82 | 89·63 | 75·82 | 74·98 | 69·58 | 87·84 | 80·37 | 72·78 |
| | SD | (8·92) | (26·42) | (28·11) | (19·24) | (17·28) | (15·35) | (34·78) | (15·83) |
| Difference | Mean | **0·65** | **5·17** | **7·91** | **9·87** | **3·65** | **-0·73** | **8·56** | **3·13** |
| | p-value | 0·192 | <0·001 | <0·001 | <0·001 | 0·001 | 0·437 | <0·001 | 0·002 |
| **Age 40-64** | | | | | | | | | |
| Falun Gong Practitioners | N | 759 | 756 | 759 | 755 | 759 | 762 | 748 | 755 |
| | Mean | 93·36 | 90·42 | 80·93 | 80·74 | 75·19 | 87·17 | 91·80 | 78·66 |
| | SD | (12·35) | (22·89) | (18·38) | (18·98) | (17·04) | (13·13) | (22·91) | (15·28) |
| Norm* | Mean | 91·00 | 81·42 | 70·21 | 66·47 | 67·17 | 87·12 | 81·54 | 73·39 |
| | SD | (15·42) | (35·06) | (28·83) | (22·21) | (19·32) | (17·03) | (35·12) | (17·37) |
| Difference | Mean | **2·36** | **9·00** | **10·72** | **14·27** | **8·02** | **0·05** | **10·26** | **5·27** |
| | p-value | <0·001 | <0·001 | <0·001 | <0·001 | <0·001 | 0·913 | <0·001 | <0·001 |
| **Age 65+** | | | | | | | | | |
| Falun Gong Practitioners | N | 181 | 178 | 180 | 173 | 175 | 180 | 176 | 173 |
| | Mean | 83·76 | 78·04 | 78·40 | 78·36 | 72·79 | 85·97 | 83·52 | 79·05 |
| | SD | (18·90) | (34·52) | (19·58) | (17·70) | (17·76) | (15·95) | (32·46) | (14·86) |
| Norm* | Mean | 69·76 | 56·71 | 61·33 | 54·58 | 58·70 | 79·05 | 68·71 | 71·19 |
| | SD | (26·99) | (46·03) | (27·87) | (23·23) | (20·47) | (23·49) | (43·40) | (18·26) |
| Difference | Mean | **14·00** | **21·33** | **17·07** | **23·78** | **14·09** | **6·92** | **14·81** | **7·86** |
| | p-value | <0·001 | <0·001 | <0·001 | <0·001 | <0·001 | <0·001 | <0·001 | <0·001 |

SF-36 domain PF= Physical function; RP= Role function; BP=Bodily pain; GH= General health; VT=Vitality and energy; SF=Social function; RE= Role function due to emotional limitations; MH=Mental health
SD= standard deviation; Difference= mean score of practitioners minus that of norm.
* Based on data from the 2001 National Health Interview Survey, Taiwan (ref 15)

**Table 4**. Effect of the length of Falun Gong practice on SF-36 domain scores among study subjects

| *Domain* | *Score* | *PF* | *RP* | *BP* | *GH* | *VT* | *SF* | *RE* | *MH* |
|---|---|---|---|---|---|---|---|---|---|
| <½ year | Mean | 87·70 | 85·96 | 74·15 | 69·42 | 65·70 | 81·53 | 88·59 | 70·81 |
| | SD | (19·76) | 27·58 | 20·42 | 22·29 | 18·64 | 16·22 | 28·21 | 17·08 |
| ½ to 1 year | Mean | 91·99 | 87·95 | 78·59 | 76·12 | 70·63 | 86·58 | 86·43 | 75·09 |
| | SD | 12·77 | 26·40 | 20·74 | 20·19 | 18·34 | 14·65 | 28·09 | 16·87 |
| 1 to 2 years | Mean | 93·48 | 90·89 | 81·21 | 82·04 | 74·42 | 86·59 | 90·58 | 77·71 |
| | SD | 12·18 | 22·08 | 18·30 | 17·75 | 16·86 | 13·68 | 24·43 | 15·12 |
| >2 years | Mean | 94·29 | 90·47 | 84·38 | 86·21 | 78·38 | 89·17 | 91·24 | 81·69 |
| | SD | 11·65 | 23·82 | 16·88 | 15·06 | 15·06 | 12·71 | 23·55 | 13·55 |
| Total* | Mean | 92·87 | 89·65 | 81·26 | 81·31 | 74·44 | 87·04 | 90·00 | 78·13 |
| | SD | 13·41 | 24·26 | 18·68 | 18·60 | 17·12 | 13·98 | 25·18 | 15·45 |
| p-value | | <0·001 | 0·117 | <0·001 | <0·001 | <0·001 | <0·001 | 0·149 | <0·001 |

SF-36 domain PF= Physical function; RP= Role function; BP=Bodily pain; GH= General health; VT=Vitality and energy; SF=Social function; RE= Role function due to emotional limitations; MH=Mental health

*Subjects who did not report length of practice (n=14, 1·2%) were excluded from this analysis·

**Table 5**· Factors influencing SF-36 domain scores and frequency of medical visits among study subjects

| Dependent Variable | *PF* | *RP* | *BP* | *GH* | *VT* | *SF* | *RE* | *MH* | Med Visits |
|---|---|---|---|---|---|---|---|---|---|
| Valid N | 1020 | 1017 | 1017 | 1011 | 1017 | 1020 | 1008 | 1013 | 1020 |
| R-squared | 0·2062 | 0·1142 | 0·1356 | 0·1981 | 0·1352 | 0·1117 | 0·0735 | 0·1343 | 0·2621 |
| Adj R-squared | 0·1879 | 0·0937 | 0·1156 | 0·1795 | 0·1152 | 0·0911 | 0·0519 | 0·1142 | 0·2451 |
| Explanation Variables | | | | | | | | | |
| Constant | 102·281*** | 105·539*** | 76·901*** | 81·525*** | 67·213*** | 90·723*** | 105·036*** | 63·928*** | 5·545** |
| Length of practice (years) | 0·457 | 0·674 | 0·884* | 2·015*** | 1·415*** | 0·715* | 0·569 | 1·060*** | -1·077*** |
| Times of reading *Zhuan Falun* | 0·634*** | 0·546** | 0·806*** | 1·117*** | 1·014*** | 0·637*** | 0·422* | 0·876*** | -0·581*** |
| Prior chronic condition (yes/no) | -2·907 *** | -4·658** | -6·439*** | -7·463*** | -5·046*** | -3·978*** | -2·621 | -4·149*** | 2·281*** |
| Female (yes/no) | -2·020** | -2·128 | -1·867 | 0·599 | -1·849 | -2·408** | -2·421 | -2·506** | -0·414 |
| Age (years) | -0·280*** | -0·366*** | -0·120 | -0·153* | 0·017 | -0·087 | -0·159 | 0·071 | 0·120*** |
| City (yes/no) | 1·273 | 0·991 | 0·327 | 2·340* | 0·572 | -0·156 | 1·269 | 1·407 | -0·363 |
| Town (yes/no) | -0·210 | -2·415 | -1·424 | 0·472 | -0·093 | -3·284*** | -1·643 | 0·493 | -0·800 |
| Household Income ($NTD/month) | 0·118 | 0·854*** | 0·432** | 0·356** | 0·453*** | 0·384*** | 0·586** | 0·208 | -0·128* |
| Education- Junior high (yes/no) | 1·798 | -3·325 | -0·297 | 0·856 | 0·011 | -2·925* | -8·484*** | 2·199 | -0·369 |
| Education- Senior high (yes/no) | 0·100 | -4·502* | 1·427 | -1·397 | -2·513 | -2·701** | -8·098*** | -1·448 | 1·200 |
| Education- College (yes/no) | 0·700 | -6·604** | 2·577 | -1·389 | -4·057** | -4·174*** | -10·131*** | -1·541 | 0·900 |
| Married (yes/no) | 0·514 | 1·487 | 1·117 | -0·057 | 0·386 | -0·863 | -3·731 | 4·846** | 1·073 |
| Divorced, widowed or separation | 0·260 | 2·615 | 0·883 | -0·652 | -0·338 | 0·560 | -3·860 | 4·312 | 0·994 |
| No child (yes/no) | 2·236 | 4·638 | 9·315*** | 1·683 | 1·479 | 1·952 | -1·699 | 5·212** | -1·573 |
| The youngest child's age 6-18 (yes/no) | 3·211** | 5·193* | 6·535*** | 1·282 | 1·146 | 2·675 | 3·916 | 2·020 | -2·094** |
| The youngest child's age ≥18 (yes/no) | 2·081 | 3·691 | 4·436 | -0·686 | -0·915 | 1·086 | 2·872 | 1·542 | -1·349 |
| Own-family work (yes/no) | -3·687*** | -7·578*** | -2·354 | -2·750 | -1·772 | -1·908 | -4·722* | -2·703 | 0·870 |
| Employer (yes/no) | -3·968* | -4·014 | 3·097 | 0·444 | 2·078 | 0·454 | -2·994 | 3·417 | -1·581 |
| Government work (yes/no) | -1·882 | -2·162 | -4·638** | -2·943 | 0·577 | -0·157 | 1·070 | -0·942 | 1·722** |
| Student (yes/no) | -4·641* | -4·875 | -0·905 | -6·430* | -4·103 | -6·918** | -26·886*** | -5·067 | 0·718 |
| Housekeeper (yes/no) | -2·292* | -3·660 | -1·333 | -3·980** | -1·027 | -0·798 | -2·540 | -0·902 | 0·828 |
| Retiree (yes/no) | -4·326*** | -8·940*** | 2·185 | -0·400 | 0·648 | 0·381 | -5·373* | -0·310 | 2·041** |
| Other job (yes/no) | 0·735 | -1·301 | 0·738 | -0·284 | 1·645 | 3·690* | -1·106 | 1·222 | -0·013 |

SF-36 domain PF= Physical function; RP= Role function; BP=Bodily pain; GH= General health; VT=Vitality and energy; SF=Social function; RE= Role function due to emotional limitations; MH=Mental health

*$p$<0·05

** $p$<0·01

*** $p$<0·001

# 台灣法輪功學員身心健康狀況問卷調查

## 民國91年10月30日

您是被隨機抽樣選中的法輪功學員。這個問卷調查的主題是「台灣法輪功學員身心健康狀況調查」，主要是希望了解您目前的健康狀況、在修煉法輪功前後健康狀況的變化以及您利用醫療保健服務的情形，作為國家促進健康政策的參考。您所提供的每一個答案，對這個調查而言，都是非常寶貴而重要的資料，所以請您一定要根據「您自己的真實情況」回答。請提供下列個人及健康相關資料。若您同意釋出您的全民健保資料與本調查資料串聯的話，請記得在問卷後面的同意書上簽名。您的個人身份的資料與個人問卷資料會被分開保存，僅以調查號互相索引，所有個人資料將被嚴格保密。謝謝您的參與和合作。

在問卷最後有一張限由停經婦女回答的附加問卷。凡是符合下面停經婦女的定義的學員請一定要填寫。停經婦女的定義：在沒有懷孕、哺乳或使用某些醫藥的情況下連續一年(十二個月)未來例假者。

您的姓名及聯絡方式;

姓名 ________________________________________

電話 ________________________________________

EMAIL ________________________________________

煉功點 ________________________________________

居住地點的郵遞區號是 □ □ □

若您不知道郵遞區號，請寫下居住的縣市和鄉鎮市區：

______________縣╲市 ________________鄉╲鎮╲市╲區

調查號 ______________________ (不用填，將自動編碼)

調查號:____________________ (不用填，將由自動號碼機打印)

## A、 <u>個人資料</u>

A1. 您的性別是 □1 男 □2 女

A2. 請問您的年齡實歲是 ____ 歲，民國 ___年出生

A3. 請問您的最高學歷是

□1 國中以下 □2 高中職 □3 專科 □4 大學 □5 研究所以上

A4. 請問您的婚姻狀況是

□1 未婚 □2 已婚 □3 同居 □4 已離婚或分居

□5 配偶去世 □6 其他 _______（請說明）

A5. 請問您的子女狀況是

□1 目前尚無子女

□2 最小的子女在 6 歲以下

□3 最小的子女在 6 至 18 歲之間

□4 最小的子女超過 18 歲，但和您同住

□5 子女都已獨立居住

□6 其他 _____________（請說明）

A6. 請問您現在的工作狀況是

□1 自雇/自己當老闆

□2 在家幫忙事業

□3 受雇替別人工作

□4 退休

□5 家庭主婦

□6 無職業

□7 其他 ______________（請說明）

A7. 請問您每月的家庭所得（如果您是單身，請問個人所得）大約是多少

□ 2 萬元及以下 □ 20,001 ~ 4 萬元 □ 40,001 ~ 6 萬元

□ 60,001 ~ 8 萬元 □ 80,001 ~10 萬元 □ 100,001~12 萬元

□ 120,001~14 萬元 □ 140,001~16 萬元 □ 16 萬元以上

## **B、 自覺健康狀態（SF-36）**

在本部份所指過去一個月內，係指從今天往前算三十天內。

B1. 一般來說，您認為您目前的健康狀況是：

□ 1 極好的 □2 很好 □3 好 □4 普通 □5 不好

B2. 和一年前比較，您認為您目前的健康狀況是？

□1 比一年前好很多 □2 比一年前好一些 □3 和一年前差不多

□4 比一年前差一些 □5 比一年前差很多

B3. 下面是一些您日常可能從事的活動，請問您目前健康狀況會不會限制您從事這 些活動?如果會，到底限制有多少?

| 活　　動 | 會，受到很多限制 | 會，受到一些限制 | 不會，完全不受限制 |
|---|---|---|---|
| a. **費力活動**，例如跑步、提重物、參與劇烈運動 | 1 | 2 | 3 |
| b. **中等程度活動**，例如搬桌子、拖地板、打保齡球、或打太極拳 | 1 | 2 | 3 |
| c. 提起或攜帶食品雜貨 | 1 | 2 | 3 |
| d. 爬**數**層樓樓梯 | 1 | 2 | 3 |
| e. 爬一層樓樓梯 | 1 | 2 | 3 |
| f. 彎腰、跪下或蹲下 | 1 | 2 | 3 |
| g. 走路**超過 1** 公里 | 1 | 2 | 3 |
| h. 走過**數個街口** | 1 | 2 | 3 |
| i. 走過**一個街口** | 1 | 2 | 3 |
| j. 自己洗澡或穿衣 | 1 | 2 | 3 |

B4. 在過去一個月內，您是否曾因爲身體健康問題，而在工作上或其它日常活動方面有下列任何的問題?

| | 是 | 否 |
|---|---|---|
| a. 做工作或其它活動的**時間**減少 | 1 | 2 |
| b. 完成的工作量比您想要**完成的較少** | 1 | 2 |
| c. 可以做的工作或其它活動的**種類**受到限制 | 1 | 2 |
| d. 做工作或其它活動**有困難** (例如，須更吃力) | 1 | 2 |

B5. 在過去一個月內，您是否曾因爲情緒問題(例如，感覺沮喪或焦慮)，而在工作上或其它日常活動方面有下列的問題?

| | 是 | 否 |
|---|---|---|
| a. 做工作或其它活動的**時間**減少 | 1 | 2 |
| b. 完成的工作量比您想要**完成的較少** | 1 | 2 |
| c. 做工作或其它活動時不如以往**小心** | 1 | 2 |

B6. 在過去一個月內，您的身體健康或情緒問題，對您與家人或朋友、鄰居、社團間的平常活動的妨礙程度如何？

☐1 完全沒有妨礙 ☐2 有一點妨礙 ☐3 中度妨礙 ☐4 相當多妨礙
☐ 5 妨礙到極點

B7. 在過去一個月內，您身體疼痛程度有多嚴重？

☐1 完全不痛 ☐2 非常輕微的痛 ☐3 輕微的痛 ☐4 中度的痛
☐5 嚴重的痛 ☐6 非常嚴重的痛

B8. 在過去一個月內，身體疼痛對您的日常工作(包括上班及家務)妨礙程度如何？

☐1 完全沒有妨礙 ☐2 有一點妨礙 ☐3 中度妨礙 ☐4 相當多妨礙
☐5 妨礙到極點

B9. 下列各項問題是關於過去一個月內您的感覺及您對周遭生活的感受，請針對每一問題選一最接近您感覺的答案。在過去一個月中有多少時候…….

| 在過去一個月中有多少時候： | 一直都是 | 大部分時間 | 經常 | 有時 | 很少 | 從不 |
|---|---|---|---|---|---|---|
| a. 您覺得充滿活力？ | 1 | 2 | 3 | 4 | 5 | 6 |
| b. 您是一個非常緊張的人？ | 1 | 2 | 3 | 4 | 5 | 6 |
| c. 您覺得非常沮喪，沒有任何事情可以讓您高興起來？ | 1 | 2 | 3 | 4 | 5 | 6 |
| d. 您覺得心情平靜？ | 1 | 2 | 3 | 4 | 5 | 6 |
| e. 您精力充沛？ | 1 | 2 | 3 | 4 | 5 | 6 |
| E. 您覺得悶悶不樂和憂鬱？ | 1 | 2 | 3 | 4 | 5 | 6 |
| g. 您覺得筋疲力竭？ | 1 | 2 | 3 | 4 | 5 | 6 |
| h. 您是一個快樂的人？ | 1 | 2 | 3 | 4 | 5 | 6 |
| i. 您覺得累？ | 1 | 2 | 3 | 4 | 5 | 6 |

B10. 在過去一個月內，您的身體健康或情緒問題有多少時候會妨礙您的社交活動（如拜訪親友等）？

□1 一直都會

□2 大部分時間會

□3 有時候會

□4 很少會

□5 從不會

B11. 下列各個陳述對您來說有多正確？

| | 完全正確 | 大部分正確 | 不知道 | 部分不正確 | 完全不正確 |
|---|---|---|---|---|---|
| a. 我好像比別人較容易生病 | 1 | 2 | 3 | 4 | 5 |
| b. 和任何一個我認識的人來比， 我和他們一樣健康。 | 1 | 2 | 3 | 4 | 5 |
| c. 我想我的健康會越來越壞 | 1 | 2 | 3 | 4 | 5 |
| d. 我的健康狀況好得很 | 1 | 2 | 3 | 4 | 5 |

## C、 煉功前後個人身心健康狀態的變化

**C1.** 整體來說，煉功前您滿意自己的**健康**嗎？
☐1極不滿意　☐2不滿意　☐3中等程度滿意
☐4滿意　☐5極滿意

C2. 整體來說，目前您滿意自己的健康嗎？
☐1極不滿意　☐2不滿意　☐3中等程度滿意
☐4滿意　☐5極滿意

C3. 煉功前，您對自己**從事日常活動的能力**滿意嗎？
☐1極不滿意　☐2不滿意　☐3中等程度滿意
☐4滿意　☐5極滿意

C4. 目前，您對自己從事日常活動的能力滿意嗎？
☐1極不滿意　☐2不滿意　☐3中等程度滿意
☐4滿意　☐5極滿意

C5. 煉功前，您常有**負面的感受**嗎？（如傷心、緊張、焦慮、憂鬱等）
☐1從來沒有　☐2不常有　☐3一半有一半沒有
☐4很常有　☐5一直都有

C6. 目前，您常有負面的感受嗎？（如傷心、緊張、焦慮、憂鬱等）
☐1從來沒有　☐2不常有　☐3一半有一半沒有
☐4很常有　☐5一直都有

D、 請問您在<u>煉功前</u>有沒有下列這些病情或症狀？若有、請比較煉功前後之病情，並且在<u>每一種疾病</u>的適當欄位上勾選(V)。若在煉功前沒有該病症、請選【沒有症狀】；若在煉功後沒有該病症、請選【沒有變化】；若在煉功後才出現該病症、請選擇適當的變化情況：

| 疾病名稱或病徵<br>【勾選(V)有病情的項目，請將該疾病名稱圈出來】 | 煉功前病情 | | | | | 煉功後變化情況 | | | | |
|---|---|---|---|---|---|---|---|---|---|---|
| | 沒有症狀 | 症狀輕微 | 症狀較嚴重 | 症狀非常嚴重 | | 完全康復 | 有所好轉 | 沒有變化 | 有所惡化 | 嚴重惡化 |
| 1、 惡性腫瘤 ，請註明部位（ ） | | | | | | | | | | |
| 2、 心臟病 | | | | | | | | | | |
| 3、 高血壓疾病 | | | | | | | | | | |
| 4、 肺部疾病（如：支氣管炎、肺氣腫、氣喘） | | | | | | | | | | |
| 5、 糖尿病 | | | | | | | | | | |
| 6、 中風（腦溢血或腦血栓） | | | | | | | | | | |
| 7、 胃潰瘍或十二指腸潰瘍 | | | | | | | | | | |
| 8、 肝臟疾病 | | | | | | | | | | |
| 9、 腎臟疾病 | | | | | | | | | | |
| 10、失眠症 | | | | | | | | | | |
| 11、攝護腺(前列腺)疾病【限男性回答】 | | | | | | | | | | |
| 12、子宮卵巢疾病【限女性回答】 | | | | | | | | | | |
| 13、紅斑性狼瘡症 | | | | | | | | | | |
| 14、甲狀腺機能亢進/減退 | | | | | | | | | | |
| 15、痛風、關節炎或風濕症 | | | | | | | | | | |
| 16、精神官能症 請註明（ ） | | | | | | | | | | |
| 17、您是不是有其他上面沒提到的病或症狀？<br>請註明（ ） | | | | | | | | | | |

## E、 醫療使用情形

E1. 請問您的身高____________公分 ，體重____________公斤

E2. 您在煉功前的那一年內，使用健保卡到第幾卡？

☐ 那一年沒有用過健保卡　　☐ E 卡 （25～30 次）
☐ A 卡（1～6 次）　　☐ D 卡 （19～24 次）
☐ B 卡（6～12 次）　　☐ F 卡或更多（31 次或更多）
☐ C 卡 （13～18 次）

E3. 您今年到現在爲止，已經使用健保卡到第幾卡？

☐ 今年沒有用過健保卡　　☐ E 卡 （25～30 次）
☐ A 卡（1～6 次）　　☐ D 卡 （19～24 次）
☐ B 卡（6～12 次）　　☐ F 卡或更多（31 次或更多）
☐ C 卡 （13～18 次）

E4. 如果看病次數下降，請說明原因 （可以複選）

☐1 身心健康，沒有需要看醫師　　☐5 家人沒有空帶你去看病
☐2 醫院或診所太遠，或交通不便　　☐6 就算去看醫生也沒有用
☐3 等候掛號或看診的時間太久　　☐7 負擔不起
☐4 工作或家事太忙，沒有時間　　☐8 其他，請說明____________

## F、 個人生活習慣在煉功前後的變化

**F1**. 請問您在煉功前是否有喝酒的習慣？

☐1 有 (續填)　　☐2 沒有（跳答 F3）

**F2.** 請問您目前(煉功後) 是否還喝酒？

☐1 已經戒酒，煉功_____個月後戒掉？　☐2 偶而少量的喝　☐3 仍然常喝

**F3**. 請問您在煉功前是否有**抽菸**習慣？

☐1 有 (續填)　　☐ 2 沒有（跳答 F5）

**F4.** 請問你目前(煉功後) 是否還**抽菸?**

☐1 已經戒菸，煉功_____個月後戒掉？　☐2 偶而還抽　☐ (3) 仍然常抽

**F5.** 請問您在<u>煉功前</u>是否有**嚼食檳榔**習慣？

□1 有 (續填)　　　□2 沒有（跳答 F7）

**F6**. 請問你<u>目前</u>(煉功後) 是否還有**嚼食檳榔**?

□1 已經戒掉，煉功____個月後戒掉？ □2 偶而還有嚼食 □3 仍然常嚼

**F7.** 請問您在<u>煉功前</u>是否**有賭博**習慣？

□1 有 (續填)　　　□2 沒有（跳答 G1）

**F8.** 請問你<u>目前</u>(煉功後) 是否還有在**賭博**?

□1 已經戒賭 ，煉功____個月後戒掉？ □2 偶而賭一下 □3 仍然常賭

## G. 修煉情形

G1. 請問您已煉功多長時間？　【您於民國　　年　　月得法】

□ 未滿半年　□ 0.5~1年　□ 1~2年　□ 2~3年

□ 3~4年　□ 4~5年　□ 5年以上 ( ___年)

G2. 請問您<u>通讀整本</u>【轉法輪】幾次？

□ 未滿 1次　□ 1～3次　□ 4～6次　□ 7～9次

□ 10次以上

G3. 請問您<u>平均</u>每天**學法時間**(大約)：　□ 未滿半小時　□ 0.5~1小時

□ 1~2小時　□ 2~3小時　□ 3小時以上 (_小時)

G4. 請問您<u>平均</u>每天**煉功時間**(大約)：　□ 未滿半小時　□ 0.5~1.5小時

□ 1.5~2.5小時　□ 2.5小時以上

G5. 請問您是否經常參加**讀書會或交流**？ □1 是　□2 否

G6. 請問您是否經常參加**集體煉功**？ □1 是　□2 否

**問卷在此結束。請繼續填寫您的全民健保資料使用<u>同意書</u>，謝謝您的合作!**

**停經女性學員請續填附加問卷！**

# 同意書

我同意將我的「全民健康保險」的所有申報資料與「台灣法輪功學身心健康狀況問卷調查」的資料串連起來，作為學術研究之用。

身份證號碼：□□□□□□□□□□

姓名：______________________________

簽名：_________________________________

日期：________________年_______________月_______________日

調查號:__________________________ (不用填，將由自動號碼機打印)

調查號 ____________________

## H. 停經婦女附加問卷（限由符合下面停經婦女定義的學員填答）

停經婦女的定義：在沒有懷孕、哺乳或使用某些醫藥的情況下連續一年(十二個月)未來例假者。

H1. 請問您何時開始停經（最後一次例假日期）？_____年_____月

H2. 煉功前，您的皮膚狀況是 □1 非常細膩光滑 □2 尚稱光滑 □3 普通 □4 有些粗糙 □5 非常粗糙

H3. 煉功後，您的皮膚狀況是 □1 非常細膩光滑 □2 尚稱光滑 □3 普通 □4 有些粗糙 □5 非常粗糙

H4. 煉功後是否重來例假？ □1 是 □2 不是 （跳答6）

H5. 煉功後多久重來例假？ □ 一週之內 □ 一週以上至一個月之內 □ 一個月以上至半年之內 □ 半年以上至一年之內 □ 一年以上

H6. 煉功前有沒有使用荷爾蒙的治療？ □1 有 □2 沒有

H7. 煉功後有沒有使用荷爾蒙的治療？ □1 有 □2 沒有

H8. 請問您是否曾做過下面的檢驗？

| | 激素水平 | 骨密度 | 婦科超音波 | 婦科體檢 | 婦科病理 |
|---|---|---|---|---|---|
| 若有、請打勾 | | | | | |

H9. 請您填寫下面表格（婦女更年期症狀自我評估表格）

其中只有熱潮紅分五等級 0 沒有；1 少于每周一次；2 每周都有；3 每天都有；4 每天多次

其餘症狀分四等級： 0 -沒有 1- 輕微 2- 中等 3-嚴重

| | | a 熱潮紅 | b 頭昏眼花 | c 頭痛 | d 暴燥 | e 情緒抑鬱 | f 失落感覺 | g 精神緊張 | h 失眠 | i 異常疲倦 | j 背痛 | k 關節疼痛 | l 肌肉酸痛 | m 面毛增多 | n 皮膚異常乾燥 | o 性慾減低 | p 性感受度降低 | q 陰道乾燥 | r 行房時感痛楚 | 總積分 |
|---|---|---|---|---|---|---|---|---|---|---|---|---|---|---|---|---|---|---|---|---|
| 煉功前 | | | | | | | | | | | | | | | | | | | | |
| 煉功後 | | | | | | | | | | | | | | | | | | | | |

## Appendix 2. Statistical Analyses

### *Power analysis*

The Taiwan norm for the SF-36 general health perception (GH) dimension was 70·42 (standard deviation (SD) 21·78).[16] We hypothesized that a 2-point change in GH scores would be meaningful in practice. To detect a 2-point change using a one-sample t-test with two-sided alpha of 0·05, 1,000 subjects will yield approximately statistical power of 82%. Assuming that the non-response (or invalid response) rate would be <30%, we planned to recruit 1,430 Falun Gong practitioners.

### *Data Analysis*

The statistical analysis was first performed using SPSS software, version 13 (SPSS Inc., Chicago, Illinois) by the PI. After the PI passed away in 2006, some analysis was checked by using R version 3.6.2 (2019), https://www.R-project.org/. All analyses were two-sided and considered statistically significant at the $p<0·05$ level. The survey data were analyzed in the following four categories:

### *1. The effect on health status compared to the 2001 Taiwan norm*

One-sample t-tests were used to compare mean scores of eight SF-36 domain scores between the study cohort of practitioners and the 2001 Taiwan norm. [14]

### *2. The effect on health status by one-way ANOVA and multivariate linear regression analyses*

One-way analysis of variance (ANOVA) was used to test whether the mean SF-36 domain scores vary among the study subjects with different lengths of time practicing Falun Gong. In addition to univariate analysis, the association among the length of practice, number of times of reading

*Zhuan Falun*, prior chronic condition, socioeconomic variables, and each of the eight SF-36 domain outcomes (dependent variables (DV)) was estimated using multiple linear regression analysis. All models were adjusted for the same list of independent or explanatory variables (IVs). The linear regression models were as follows:

$$Score_j=\beta_{0j} +\beta_{1j}\ \textit{length of practice} +\beta_{2j}\ \textit{number of times reading Zhuan Falun}$$
$$+\beta_{3j}\ \textit{prior chronic conditions} +\beta_{4j}^{T}\ \textit{socioeconomic characteristics} + \varepsilon_j,$$

where $j$ =PF, RP, BP, GH, VT, SF, RE, and MH are the eight SF-36 domain scores, and the socioeconomic variables include gender, age, residence area, monthly income, education level, marriage status, child status, and employment status (see Table 1). An asterisk sign (*) indicates the corresponding variable was a key factor, meaning that one unit change in the value of the independent variable has a significant ($p<0{\cdot}05$) impact on the dependent variable when the values of other independent variables hold constant (i.e., being controlled).

***3. The effect on the number of medical and preventative care visits***

Similarly, the same linear regression analysis was used to examine the effect of practicing Falun Gong on the number of medical visits (as a DV). This part of analysis provided objective assessment of the effect of practicing Falun Gong on the medical facility utilization, which may also be treated as another proxy of the individual's health status.

***4. Testing interaction effect due to prior chronic conditions in multiple linear regression models***

The null hypothesis was that there was no difference in the effects of practicing Falun Gong on health status between practitioners who had prior chronic conditions and those who did not have any. We tested whether the regression models differed by status of prior medical condition. The unrestricted model included the interaction terms of the status of prior conditions with all the explanatory variables. The restricted model did not include any of the interaction terms. F-tests were used to evaluate whether the whole set of interaction terms were significant.